\documentclass[preprint,12pt, 5p, twocolumn]{elsarticle}

\usepackage{amssymb}
\usepackage{caption}
\usepackage{subcaption}
\usepackage{threeparttable}

\journal{JQSRT}

\begin{document}

\begin{frontmatter}



\title{Comment on: ``Hyperfine structure measurements of Co~I and Co~II with Fourier transform spectroscopy" by Fu et al. [JQSRT 2021, 107590]}


\author[inst1]{Milan Ding}

\affiliation[inst1]{organization={Physics Department, Imperial College London},
            addressline={Prince Consort Road}, 
            city={London},
            postcode={SW7~2AZ}, 
            country={UK}}

\author[inst1]{Juliet C. Pickering}

\begin{abstract}
This comment points out errors in the analysis of 61 magnetic hyperfine structure ($A$) constants of Co~II energy levels by Fu et al. [JQSRT 2021, 107590]. The paper was published without full awareness of the extensive literature already available for Co~II hyperfine $A$ constants at the time; 57 of 58 $A$ constants that were claimed to have been measured for the first time had already been measured by the prior work of Ding \& Pickering [ApJS 2020, 251:24], who had published $A$ constants for 292 levels of Co~II. The $A$ constant of 3d$^6$4s$^2$~a$^5$D$_4$ has been determined by Fu et al. [JQSRT 2021, 107590] for the first time to be $12.0\pm1.8$~mK (1~mK~$=$~0.001~cm$^{-1}$), which was found to agree with line profiles observed by Ding \& Pickering [ApJS 2020, 251:24]. Discrepancies in 17 $A$ constants of Fu et al. [JQSRT 2021, 107590] were found, which are likely due to the analysis of weak, experimentally unclassified transitions with Ritz wavenumbers 25453.966~cm$^{-1}$ and 25149.948~cm$^{-1}$ by Fu et al. [JQSRT 2021, 107590] for the $A$ constants of the energy levels 3d$^7$($^2$G)4s~a$^3$G$_5$ and 3d$^7$($^2$P)4s~c$^3$P$_2$ respectively. Fewer transitions and poorer quality spectra analysed by Fu et al. [JQSRT 2021, 107590] are also concluded to have contributed to disagreements in the 17 $A$ constants.
\end{abstract}




\end{frontmatter}


\section{Introduction}
Fu~et~al.~(2021)~\cite{fu2021hyperfine} had recently published magnetic hyperfine interaction ($A$) constants for 61 energy levels of Co~II by analysing hyperfine structure (HFS) of spectral lines measured by Fourier transform spectroscopy. This paper was published in May 2021 apparently without the authors and reviewers being aware of the prior existing literature at the time, specifically, the substantial work of Ding \& Pickering~(2020)~\cite{ding2020measurements} published in December 2020, which reported $A$ constants for 292 Co~II energy levels. These include values for all Co~II $A$ constants published by Fu~et~al.~\cite{fu2021hyperfine}, which were incorrectly claimed to have been measured for the first time, except for the energy level 3d$^6$4s$^2$~a$^5$D$_4$.

Out of the 60 $A$ constants reported by both publications, 43 were in agreement within uncertainties. In this comment, discrepancies of the other 17 $A$ constants are discussed from the point of view of the authors of Ding \& Pickering~\cite{ding2020measurements}. The single new $A$ constant published by Fu~et~al.~\cite{fu2021hyperfine} for 3d$^6$4s$^2$~a$^5$D$_4$ is found to agree with corresponding line profiles from the visible spectra analysed by Ding \& Pickering~\cite{ding2020measurements}, using the $A$ constants of connecting levels published by both papers.

\section{Cobalt Fourier transform spectra}
Fu~et~al.~\cite{fu2021hyperfine} reported using 9 cobalt spectra from the National Solar Observatory (Kitt Peak, USA) historical archive (http://diglib.nso.edu/). However, these could not be found in the archive using the dates and serial numbers stated on table 1 of Fu~et~al.~\cite{fu2021hyperfine}. The spectrum transformation dates were likely listed instead of the dates at which the spectra were recorded. This is evident from the spectrum with index 7 on table~1 of Fu~et~al.~\cite{fu2021hyperfine} - it was dated 1994, but its transformation date, serial number and parameters match exactly with a spectrum listed on table 2 of Lawler~et~al.~(2018)~\cite{lawler2018transition}, which was in fact measured in 1983 and also used by Lawler~et~al.~\cite{lawler2018transition} for Co~II HFS analysis.

The parameters of spectra listed on table~1 of Fu~et~al.~\cite{fu2021hyperfine} indicate no spectrum in common between the set of spectra used by Fu~et~al.~\cite{fu2021hyperfine} and the set of spectra used by Ding \& Pickering~\cite{ding2020measurements} for HFS analysis. Moreover, the UV spectra analysed by Fu~et~al.~\cite{fu2021hyperfine} were expected to have higher noise compared to those analysed by Ding \& Pickering~\cite{ding2020measurements}, due to their much wider spectral ranges and fewer interferogram co-adds.

\section{Discrepancies and dependencies}
All 17 energy levels with $A$ constants in disagreement between Fu~et~al.~\cite{fu2021hyperfine} and Ding \& Pickering~\cite{ding2020measurements} are listed on table \ref{tab 1: levels}. The first three columns specify level configuration, term label and $J$ value. The fourth column lists level energies from Ding \& Pickering~\cite{ding2020measurements}. The fifth column indicates the number of lines used by Ding \& Pickering~\cite{ding2020measurements} to estimate the values and uncertainties of $A$ constants in their publication. The sixth and seventh columns are $A$ constant mean values and uncertainties $\Delta A$ determined by Ding \& Pickering~\cite{ding2020measurements}, values in the eighth and ninth columns are those determined by Fu~et~al.~\cite{fu2021hyperfine}, which were converted from MHz to mK (1~mK~$=$~0.001~cm$^{-1}$=29.979~MHz). The final column lists the difference in $A$ constants, all of which were above discrepancies allowed by uncertainties.
\begin{table*}
\renewcommand{\arraystretch}{1.1}
\centering
\caption{Table of 17 Co~II energy levels and their inconsistent $A$ constants between results from Ding \& Pickering~\cite{ding2020measurements} and Fu~et~al.~\cite{fu2021hyperfine}, 1~mK = 0.001~cm$^{-1}$.}
\label{tab 1: levels}
\begin{threeparttable}
\begin{tabular}{lllrrrrrrr}
\hline
       Config. &  Term &  $J$ &     Energy\textsuperscript{a} &     $N$ &     
       $A$\textsuperscript{a}  &  $\Delta A$\textsuperscript{a} &  $A$\textsuperscript{b} &   $\Delta A$\textsuperscript{b} &  Diff. \\
               &       &    &     (cm$^{-1}$) &  &   (mK) &  (mK)   &  (mK)  &  (mK)       &  (mK) \\
\hline
         3d$^8$ &  a$^3$P &  1 &  13404.321 &   2 &  -3.0 &    5.0 &  -11.9 &        1.6 &              8.9 \\
  3d$^7$($^2$G)4s &  a$^3$G &  5 &  21624.528 &  13 &  40.8 &    0.7 &    4.4 &        0.9 &             36.4 \\
  3d$^7$($^2$G)4s &  a$^3$G &  4 &  22009.344 &  17 &  22.7 &    0.4 &   30.8 &        5.9 &             -8.1 \\
  3d$^7$($^2$P)4s &  c$^3$P &  2 &  24886.400 &  14 &  51.8 &    1.0 &   11.2 &        3.5 &             40.6 \\
  3d$^7$($^2$P)4s &  a$^1$P &  1 &  27585.138 &   4 &  26.4 &    0.9 &   15.0 &        8.7 &             11.4 \\
  3d$^7$($^2$H)4s &  a$^3$H &  4 &  27902.161 &   4 &   9.5 &    1.9 &  -10.3 &        1.1 &             19.8 \\
  3d$^7$($^2$H)4s &  a$^1$H &  5 &  30567.173 &  10 &  29.9 &    1.3 &   -7.6 &        0.9 &             37.5 \\
  3d$^7$($^4$F)4p &  z$^3$D &  1 &  52684.634 &   6 &  46.7 &    1.2 &   40.0 &        1.1 &              6.7 \\
  3d$^7$($^4$P)4p &  y$^5$D &  3 &  61240.746 &   4 &   8.4 &    1.6 &    5.6 &        0.9 &              2.8 \\
  3d$^7$($^2$G)4p &  z$^3$H &  5 &  63306.686 &   7 &  22.1 &    0.7 &   24.5 &        0.9 &             -2.4 \\
  3d$^7$($^4$P)4p &  y$^3$D &  3 &  63586.987 &   8 &   5.6 &    1.0 &  -23.4 &        3.5 &             29.0 \\
  3d$^7$($^2$G)4p &  z$^3$H &  6 &  63597.396 &   1 &  16.0 &    2.0 &  -14.4 &        0.9 &             30.4 \\
  3d$^7$($^2$G)4p &  z$^1$G &  4 &  64401.359 &   8 &  22.2 &    1.1 &  -23.6 &        0.9 &             45.8 \\
  3d$^7$($^2$P)4p &  z$^3$P &  1 &  65028.512 &   5 &   5.1 &    0.7 &  -39.8 &        3.5 &             44.9 \\
  3d$^7$($^2$P)4p &  x$^3$D &  3 &  67524.021 &   3 &  10.5 &    1.5 &  -19.4 &        3.5 &             29.9 \\
 3d$^7$(a$^2$D)4p &  w$^3$D &  1 &  69317.077 &   1 &  22.2 &    0.7 &   -8.3 &       10.0 &             30.5 \\
  3d$^7$($^2$H)4p &  z$^1$I &  6 &  69617.495 &   1 &  24.0 &    2.0 &   -6.7 &        3.4 &             30.7 \\
\hline
\end{tabular}
\begin{tablenotes}
\item[a]{Ding \& Pickering~\cite{ding2020measurements}}
\item[b]{Fu~et~al.~\cite{fu2021hyperfine}}
\item Columns show - level configuration, term label, $J$ value, level energy from Ding \& Pickering~\cite{ding2020measurements}, the number of lines ($N$) used by Ding \& Pickering~\cite{ding2020measurements} to estimate $A$ constant mean values and uncertainties, $A$ constants and uncertainties by Ding \& Pickering~\cite{ding2020measurements} and Fu~et~al.~\cite{fu2021hyperfine}, and the difference between $A$ constants from the two publications.
\end{tablenotes}
\end{threeparttable}
\end{table*}

\begin{figure}
    \centering
    \includegraphics[width=8cm]{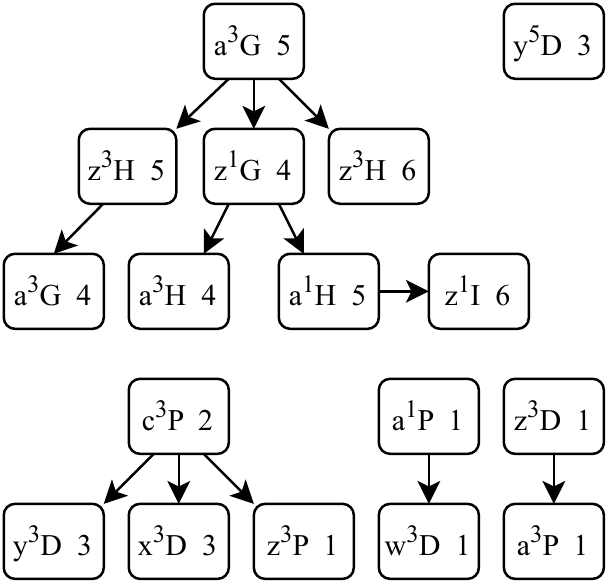}
    \caption{Tree diagrams of 17 Co~II energy levels illustrating dependencies of their $A$ constants in the analysis by Fu~et~al.~\cite{fu2021hyperfine}, as indicated by table 3 of their paper. The arrows indicate which $A$ constant would have affected the determination of another, e.g., the $A$ constant of the first energy level of each tree affected hyperfine structure analysis for all other levels on the tree.}
    \label{fig1}
\end{figure}

Figure \ref{fig1} arranges the 17 levels, labelled using term and J value, into tree diagrams showing $A$ constant dependencies during the HFS analysis by Fu~et~al.~\cite{fu2021hyperfine} according to the combining levels listed on table 3 of Fu~et~al.~\cite{fu2021hyperfine}.

\section{Experimentally unclassified transitions analysed}
All $A$ constants of energy levels on the largest tree of figure \ref{fig1} were affected by the HFS analysis by Fu~et~al.~\cite{fu2021hyperfine} for the energy level 3d$^7$($^2$G)4s~a$^3$G$_5$ at 21624.528~cm$^{-1}$. In table 3 of Fu~et~al.~\cite{fu2021hyperfine}, the transition used to determine the $A$ constant of this level was at Ritz wavenumber 25453.966~cm$^{-1}$, combining with the energy level 3d$^7$($^4$F)4p~z$^5$G$_6$ at 47078.494~cm$^{-1}$. This transition has yet to be classified experimentally (see the spectrum and term analysis of Co~II by Pickering~et~al.~\cite{pickering1998spectrum} and the observed Co~II atomic transitions from the National Institute of Standards and Techonology Atomic Spectra Database~\cite{kramida2021}). The transition has been predicted by R.~L.~Kurucz (1988)~\cite{kurucz1988} with a very weak transition probability with a log~gf of $-3.6$ (available on the Vienna Atomic Line Database \cite{ryabchikova2015major}). In the spectra analysed by Ding \& Pickering~\cite{ding2020measurements}, only noise was visible at this wavenumber.

Similarly for the second largest tree of figure \ref{fig1}, the $A$ constant of 3d$^7$($^2$P)4s~c$^3$P$_2$ at 24886.400~cm$^{-1}$ was analysed by Fu~et~al.~\cite{fu2021hyperfine} using 3d$^7$($^4$F)4p~z$^3$G$_3$ at 50036.348~cm$^{-1}$ as the combining level with the transition at Ritz wavenumber 25149.948~cm$^{-1}$. This transition is again not seen with an experimentally classified line and has a very weak predicted log~gf of $-3.9$ \cite{kurucz1988}, its signal-to-noise ratio of 2 in the spectra used by Ding \& Pickering~\cite{ding2020measurements} was too low for HFS analysis.

Within the 9 spectra analysed by Fu~et~al.~\cite{fu2021hyperfine}, the two transitions at 25453.966~cm$^{-1}$ and at 25149.948~cm$^{-1}$ were likely too noisy for meaningful hyperfine structure analysis. This would have caused cascades in error for 10 other $A$ constants on the trees of 3d$^7$($^2$G)4s~a$^3$G$_5$ and 3d$^7$($^2$P)4s~c$^3$P$_2$.

\section{Lack of validation of $A$ constants using transitions involving other levels}
Checking consistency between the $A$ constants published by Fu~et~al.~\cite{fu2021hyperfine} further supported the conclusion that their determined $A$ constants were erroneous for the levels 3d$^7$($^2$G)4s~a$^3$G$_5$ and 3d$^7$($^2$P)4s~c$^3$P$_2$.

In HFS analysis using Fourier transform spectroscopy, least-squares fitting of observed line profiles of one transition often cannot fully constrain $A$ constant values, i.e. ambiguities or a range of $A$ values can fit the spectral line, causing large uncertainties (see Ding \& Pickering~\cite{ding2020measurements} and Lawler~et~al.~\cite{lawler2018transition} for example). In general, the best practice is to analyse multiple transitions involving as many energy levels as possible for constraint and validation when estimating the $A$ constant of a single level.

\begin{figure*}
     \centering
     \begin{subfigure}{0.49\textwidth}
         \centering
         \includegraphics[width=\textwidth]{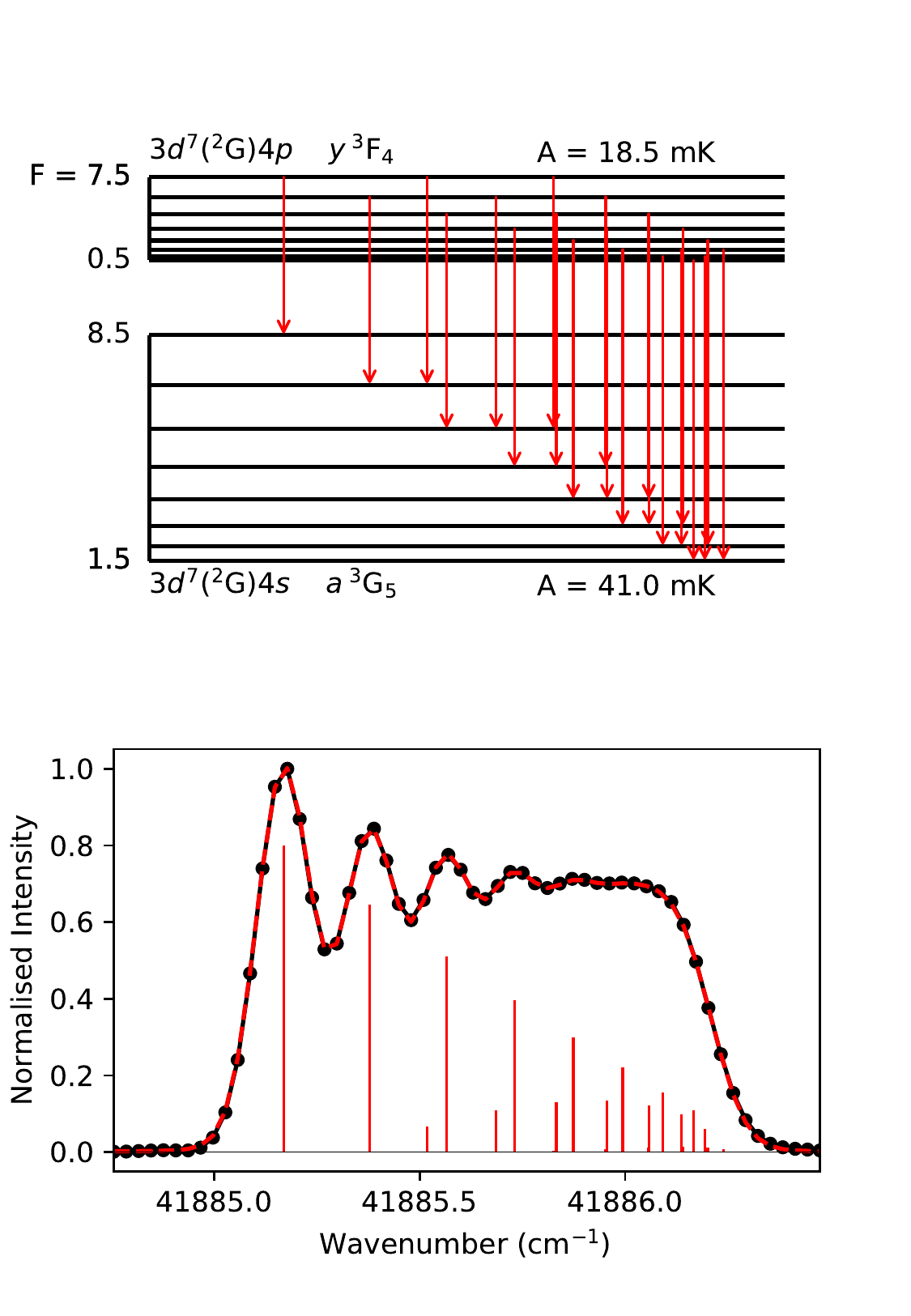}
         \caption{Spectral line fit using $A$ constants of Ding \& Pickering~\cite{ding2020measurements}.}
         \label{fig2a}
     \end{subfigure}
     \begin{subfigure}{0.49\textwidth}
         \centering
         \includegraphics[width=\textwidth]{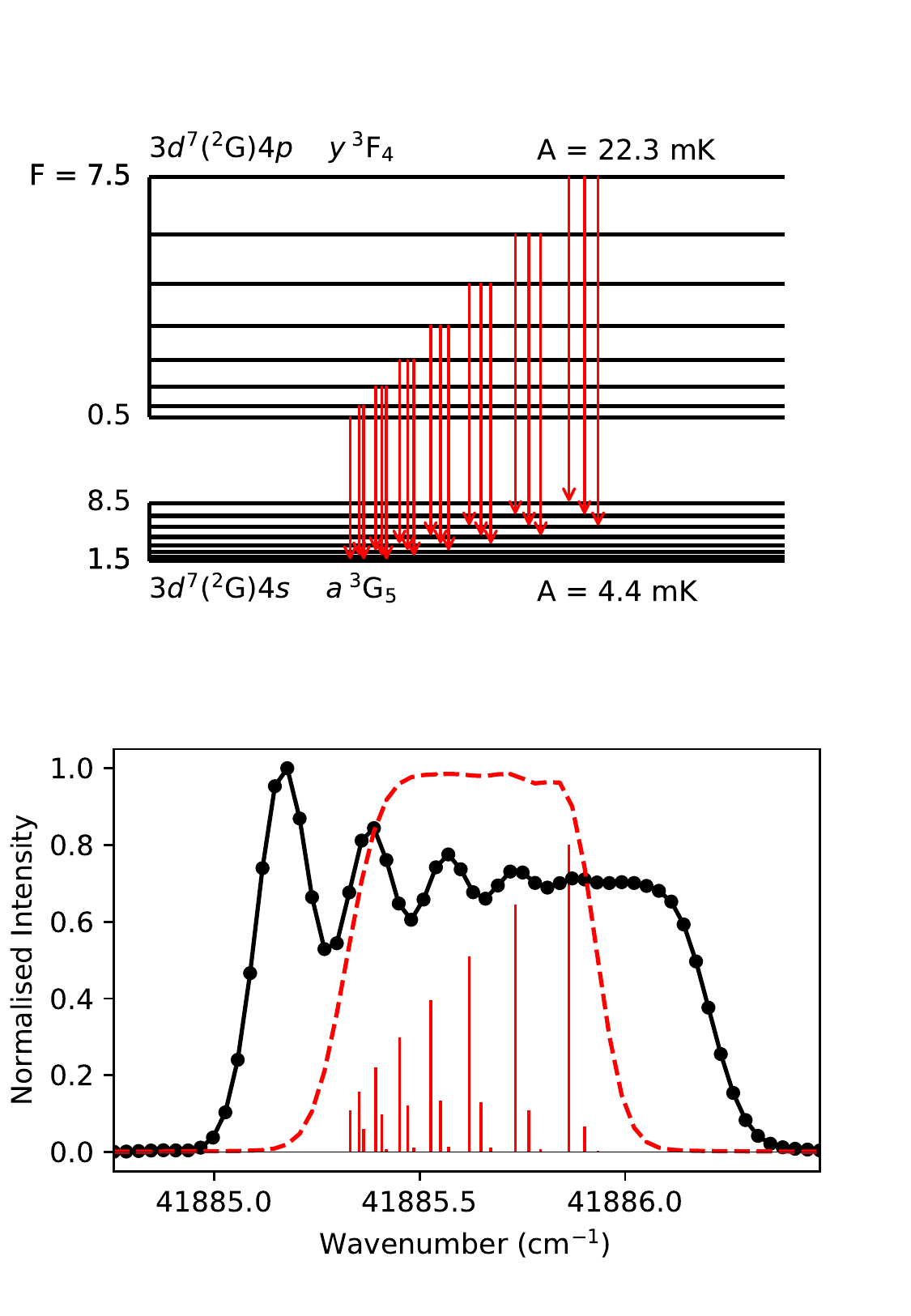}
         \caption{Spectral line fit using $A$ constants of Fu~et~al.~\cite{fu2021hyperfine}.}
         \label{fig2b}
     \end{subfigure}
     \caption{Hyperfine structure analysis of the transition between 3d$^7$($^2$G)4p~y$^3$F$_4$ and 3d$^7$($^2$G)4s~a$^3$G$_5$ of Co~II at 41885.65~cm$^{-1}$ using $A$ constants of Ding \& Pickering~\cite{ding2020measurements} (a) and Fu~et~al.~\cite{fu2021hyperfine} (b). Top - diagrams illustrating hyperfine transitions between the two levels using estimated $A$ constants (1~mK = 0.001~cm$^{-1}$), level splittings not to scale. Bottom - observation (black) and fits (red dashed) of the line in a UV spectrum analysed by Ding \& Pickering~\cite{ding2020measurements} with a signal-to-noise ratio of 950, the red vertical lines show relative intensities and wavenumbers of component transitions indicated above on the transition diagrams.}
     \label{fig2}
\end{figure*}

Fu~et~al.~\cite{fu2021hyperfine} had published $A$ constants for both levels 3d$^7$($^2$G)4p~y$^3$F$_4$ and 3d$^7$($^2$G)4s~a$^3$G$_5$.
However, these two $A$ constants do not reproduce the line profile of the transition between their corresponding levels. The transition between these two levels was observed and analysed in the UV spectra used by Ding \& Pickering~\cite{ding2020measurements}, see figure \ref{fig2}. With a high signal-to-noise ratio of 950, this transition is expected to be observed in the archival spectra analysed by Fu~et~al.~\cite{fu2021hyperfine}. In figure \ref{fig2b}, only half of the total line width was characterised by the $A$ constants from Fu~et~al.~\cite{fu2021hyperfine}, the highest relative intensity component was also on the wrong side. It is assumed that Fu~et~al.~\cite{fu2021hyperfine} did not consider this transition at 41885.65~cm$^{-1}$ during their HFS analysis, as it was not listed by table 3 nor shown in figure 1 of their paper. This fit for validation would have avoided discrepancies in 7 other $A$ constants which were dependent on the HFS analysis of 3d$^7$($^2$G)4s~a$^3$G$_5$, as indicated on its tree in figure \ref{fig1}.

\begin{figure*}
     \centering
     \begin{subfigure}{0.49\textwidth}
         \centering
         \includegraphics[width=\textwidth]{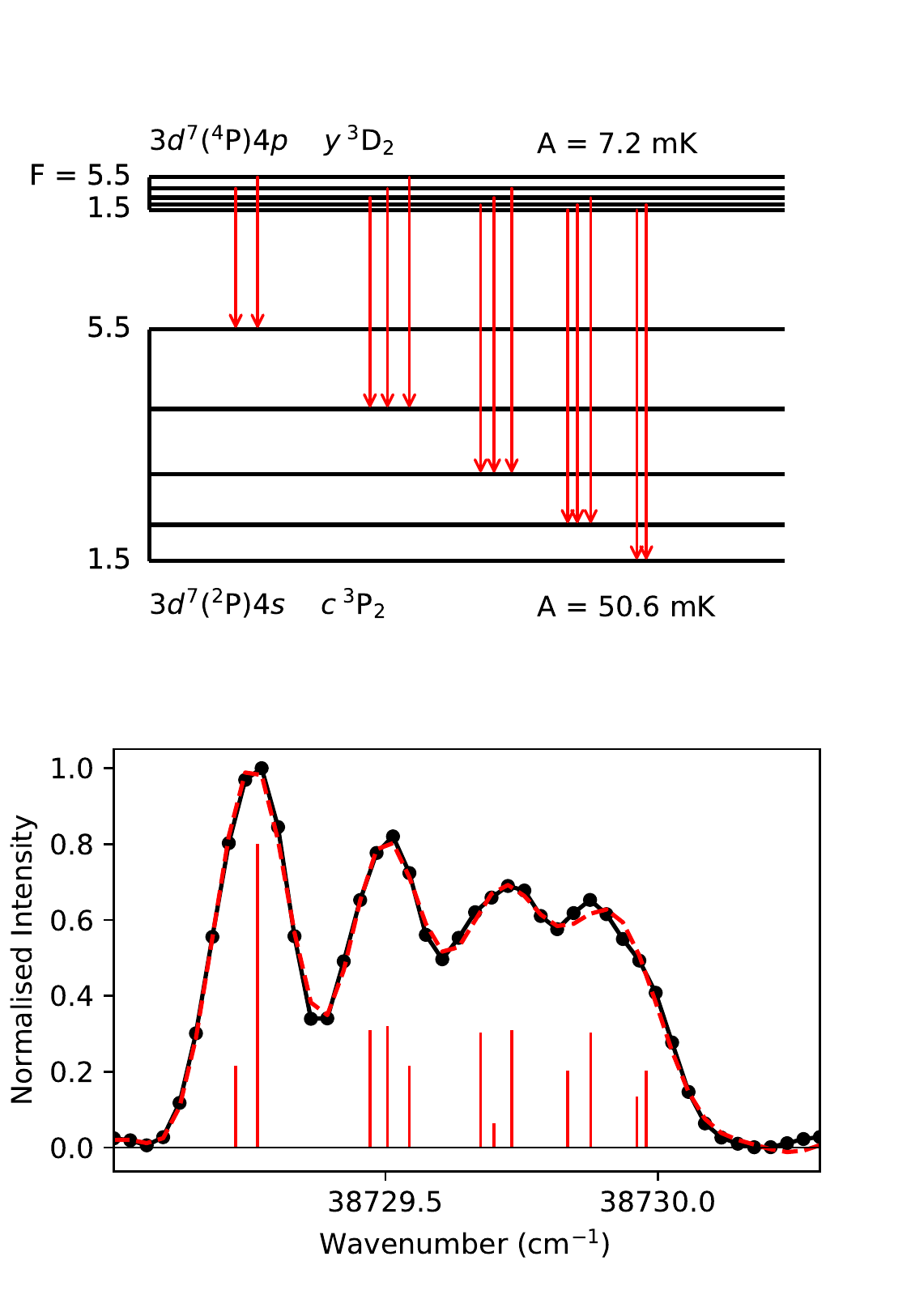}
         \caption{Spectral line fit using $A$ constants of Ding \& Pickering~\cite{ding2020measurements}.}
         \label{fig3a}
     \end{subfigure}
     \begin{subfigure}{0.49\textwidth}
         \centering
         \includegraphics[width=\textwidth]{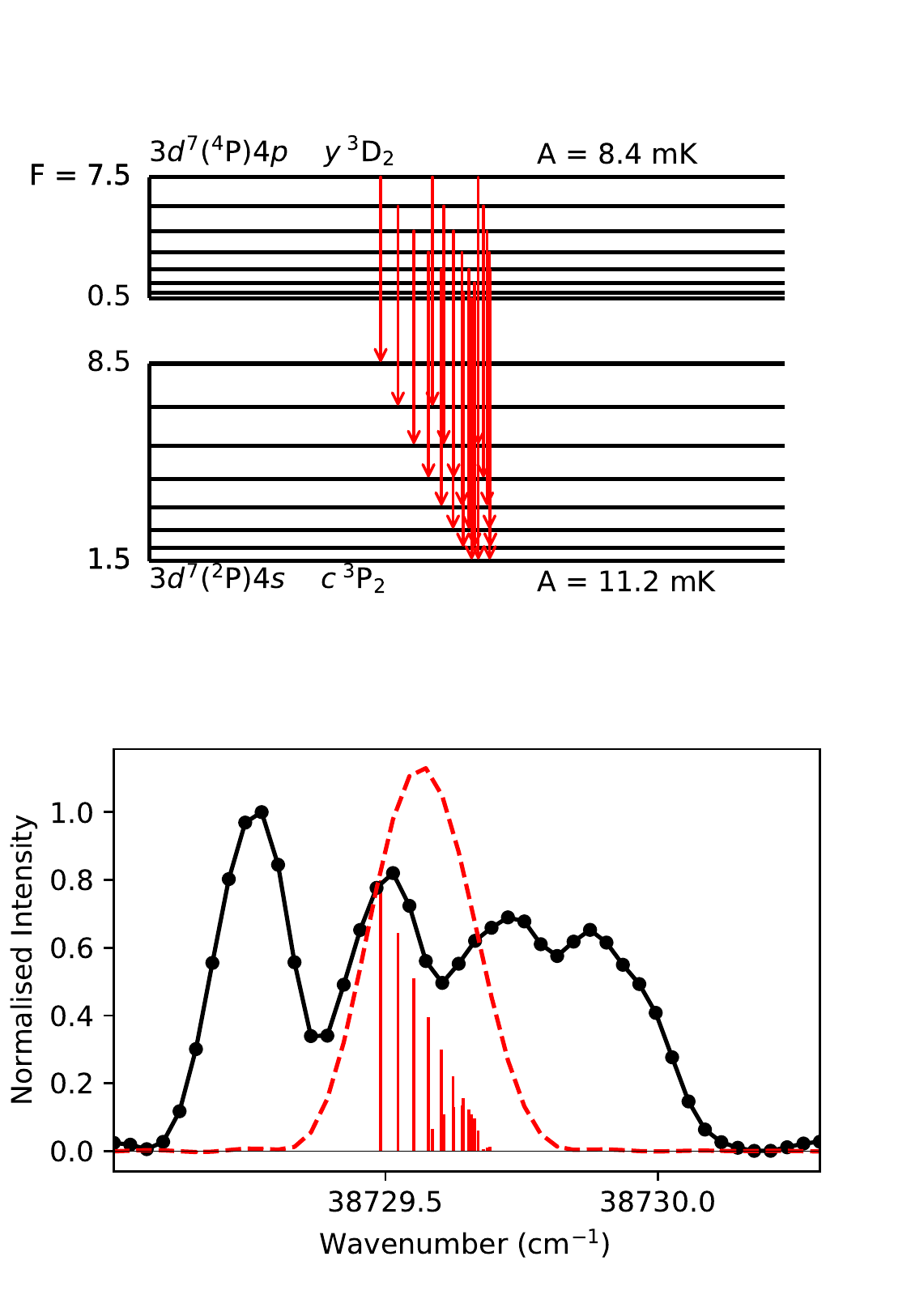}
         \caption{Spectral line fit using $A$ constants of Fu~et~al.~\cite{fu2021hyperfine}.}
         \label{fig3b}
     \end{subfigure}
     \caption{Hyperfine structure analysis of the transition between 3d$^7$($^4$P)4p~y$^3$D$_2$ and 3d$^7$($^2$P)4s~c$^3$P$_2$ of Co~II at 38729.59~cm$^{-1}$ using $A$ constants of Ding \& Pickering~\cite{ding2020measurements} (a) and Fu~et~al.~\cite{fu2021hyperfine} (b). Top - diagrams illustrating hyperfine transitions between the two levels using estimated $A$ constants (1~mK = 0.001~cm$^{-1}$), level splittings not to scale. Bottom - observation (black) and fits (red dashed) of the line in a UV spectrum analysed by Ding \& Pickering~\cite{ding2020measurements} with a signal-to-noise ratio of 40, the red vertical lines show relative intensities and wavenumbers of component transitions indicated above on the transition diagrams.}
     \label{fig3}
\end{figure*}

Likewise, the discrepancies in 3d$^7$($^2$P)4s~c$^3$P$_2$ and 3 other energy levels of the second largest tree in figure \ref{fig1} could have been avoided if the transition between energy levels 3d$^7$($^4$P)4p~y$^3$D$_2$ and 3d$^7$($^2$P)4s~c$^3$P$_2$ at 38729.59~cm$^{-1}$ was validated. In the spectra analysed by Ding \& Pickering~\cite{ding2020measurements}, this transition could not be fitted using the two corresponding $A$ constants published by Fu~et~al.~\cite{fu2021hyperfine}, see figure \ref{fig3}.

\section{Remaining discrepancies}
Discrepancies in 5 other $A$ constants reported by Fu~et~al.~\cite{fu2021hyperfine} were from their HFS analysis of energy levels 3d$^7$($^4$P)4p~y$^5$D$_3$, 3d$^7$($^2$P)4s~a$^1$P$_1$, and 3d$^7$($^4$F)4p~z$^3$D$_1$, as shown in figure \ref{fig1}. The disagreements in the $A$ constants of these 3 levels are much smaller, as they were determined by Fu~et~al.~\cite{fu2021hyperfine} using observed and classified transitions. Though not discussed in detail, further investigation suggested the discrepancies to have arisen from the poorer quality of spectra analysed by Fu~et~al.~\cite{fu2021hyperfine} compared to those analysed by Ding \& Pickering~\cite{ding2020measurements}; results from Fu~et~al.~\cite{fu2021hyperfine} were likely affected by factors such as lower signal-to-noise ratios, broader Doppler line widths, poorer resolutions and possibly higher amounts of self-absorption. 


\section{Conclusion}
Fourier transform spectroscopy offers sufficient spectral resolution for accurate atomic HFS measurements, but much care is needed during line profile fitting and analysis. For Co~II, magnetic hyperfine interaction $A$ constants for 292 energy levels were previously known, of which 264 were measured for the first time by Ding \& Pickering~\cite{ding2020measurements} using Fourier transform spectroscopy. A recent analysis of Co~II Fourier transform spectra by Fu~et~al.~\cite{fu2021hyperfine} published 61 $A$ constants without crucial awareness of this existing literature at the time, 17 of which were found to be inconsistent with results from Ding \& Pickering~\cite{ding2020measurements}. One new $A$ constant of Co~II was determined by Fu~et~al.~\cite{fu2021hyperfine} for the energy level 3d$^6$4s$^2$~a$^5$D$_4$, its value of $12.0\pm1.8$~mK was found to agree with line profiles observed by Ding \& Pickering~\cite{ding2020measurements}.

Most discrepancies are suspected to be from the HFS analysis by Fu~et~al.~\cite{fu2021hyperfine} of the experimentally unclassified transitions with Ritz wavenumbers 25453.966~cm$^{-1}$ and 25149.948~cm$^{-1}$ for the $A$ constants of the two energy levels 3d$^7$($^2$G)4s~a$^3$G$_5$ and 3d$^7$($^2$P)4s~c$^3$P$_2$ respectively. This likely caused a cascade of errors in the HFS analysis that was dependent on the $A$ constants of these two levels, as the two $A$ constants were shown to be unable to reproduce observed line profiles at 41885.65~cm$^{-1}$ and 38729.59~cm$^{-1}$. Note that Ding \& Pickering~\cite{ding2020measurements} used 13 and 14 lines to find the $A$ constants for 3d$^7$($^2$G)4s~a$^3$G$_5$ and 3d$^7$($^2$P)4s~c$^3$P$_2$ respectively. Although the spectra analysed by Fu~et~al.~\cite{fu2021hyperfine} were likely poorer in quality compared to those analysed by Ding \& Pickering~\cite{ding2020measurements}, all discrepancies would have been minimal, if Fu~et~al.~\cite{fu2021hyperfine} had analysed more transitions involving other energy levels for the 17 discrepant $A$ constants.

\section*{Acknowledgements}
This work is funded by the STFC (UK).
\bibliographystyle{elsarticle-num-names} 
\bibliography{cas-refs}







\end{document}